\documentclass[epjST]{svjour}
\usepackage{graphics}
\begin{document}
\title{Low-dimensional physics of ultracold gases with bound states and the sine-Gordon model}

\author{Thierry Jolicoeur\inst{1} \and  Evgeni Burovsky\inst{2} \and Giuliano Orso\inst{3}}
\institute{
LPTMS, CNRS and Universit\'e Paris-Sud, rue G. Cl\'emenceau,
91405 Orsay, France
\and
Department of Physics,
Lancaster University,
Lancaster,
LA1 4YB, England
\and
Universit\'e Paris Diderot-Paris7,
Laboratoire Mat\'eriaux et Ph\'enom\`enes Quantiques,
Batiment Condorcet,
75205 Paris, France}
\abstract{
One-dimensional systems of interacting atoms are an ideal laboratory to study
the Kosterlitz-Thouless phase transition. In the renormalization group picture
there is essentially a two-parameter phase diagram to explore. We first present
how detailed experiments have shown direct evidence for the theoretical treatment
of this transition. Then generalization to the case of two-component systems with bound
state formation is discussed. Trimer formation in the asymmetric attractive Hubbard model
involve in a crucial this kind of physics. } 
\maketitle
\section{Introduction}
\label{intro}

  It is now routinely feasible to create one-dimensional ultracold quantum gases either Bose
or Fermi~\cite{PitaevskiiRMP2008}. With a tube-shape potential the kinetic energy is totally frozen in the direction perpendicular to
the tube. It should be noted that this by far the best practical way to achieve low-dimensionality.
Another physical system that has been the focus of many studies is the two-dimensional electron gas.
In this case the electronic motion perpendicular to the plane is quantized and one hopes that the characteristic
energy of excitation of the perpendicular motion is higher than the other physically relevant energy scales.
However even if this may be achieved, one should keep in mind that virtual transitions to the excited states
renormalize the interactions in a way which is difficult to control. As such, the study of gases in confining tubes
is cleaner and allows more detailed and precise comparisons with theory.

The low-dimension world has enhanced fluctuations that may forbid long-range order.
This is well known in the context of magnetic systems where the Mermin-Wagner theorem
gives restrictions of the allowed ordering and the corresponding phase transitions.
One-dimensional systems at zero temperature are allowed to possess order corresponding
to discrete symmetry breaking and there may be quantum phase transitions with appearance
of Ising order. For example the XXZ antiferromagnetic spin-1/2 chain has a N\'eel ordered phase
when the anisotropy is larger than one. This phase has a true discrete long-range order at T=0K
which is allowed by the Mermin-Wagner theorem. When the anisotropy is less than one i.e.
of XY type, the mean-field like XY long-range order is replaced by algebraic correlations decaying
to zero at infinity. The appearance of phases with quasi long-range order is the striking feature
of one dimensional quantum systems. The quantum phase transition that takes place at T=0K
between N\'eel and algebraic phases is of the Kosterlitz-Thouless (KT) type, an infinite-order phase transition.
In the realm of quantum gases we also expect to observe the KT transition in various guises.
This brief review presents recent works that are directly related to this special phenomenon.
After a brief presentation of theoretical underpinnings in section~\ref{theory}, we discuss
direct evidence for the control of the renormalization group phase diagram pf the KT transition
in section~\ref{KT}. We next discuss the more involved system with two-component gases
and the possibility of formation of many-particle bound states in section~\ref{trimers}.
Finally we present we give some conclusion about future prospects.

%

\section{Field theory for low-dimensional gases}
\label{theory}

  If we consider quantum systems in two or three space dimensions, then we classify excitations above the ground states
in terms of particle-like excitations as well as collective excitations. In the 
case of Fermi systems there is a sharp Fermi surface at zero temperature and we can make
low-energy particle-hole
excitations above it. It is well known that there are also sound-like modes that can be excited.
The particle or holes are in general dressed by interactions and have a complicated
wavefunction. When there is still adiabatic continuity with respect to the non-interacting limit
we have the Landau liquid description of the system which is relevant.
Eventually the system may develop a superfluid/superconducting instability and
the ensuing quasiparticles are gapped. But we still have a description in terms
of Bogolyubov gapped quasiparticles and collectives with mutual coupling.
This picture is successful in various condensed matter systems as well as in
the description of atomic nuclei. Bose systems behave similarly~:
we can have individual particles excited above the condensates as well as
various sound modes that are collective excitations of the system as a whole.
This is a reasonably accurate description of the various liquid Heliums
for example. If interactions are very strong the boundary between these entities
particles and collective modes may become blurred and not so useful.

  The high-dimension picture above is drastically modified in one space dimension.
Here indeed all excitations are collective in character. 
The reason for this striking phenomenon is the change of available phase space.
As a consequence all excitations can be expressed in terms of a single continuous
Bose quantum field. This is the basis of the bosonization technique.
It is important to note that the Bose field does not have the same physical significance as in 
quantum field field theory in usual 3+1 dimensional Minkovsky space. Indeed the creation
operator in second-quantized formalism is written as~:
\begin{equation}
\Psi^\dagger (x) \sim \Big(\rho - \frac{1}{\pi} \partial_x \phi
\Big)^{1/2} \sum_{n} e^{ i n(k x -
\displaystyle{\phi})} \, e^{-i\displaystyle{\theta}}\,,
\label{defphi}
\end{equation}
where $k=\pi\rho$, $\rho$ being the average density, $\phi$ is a Bose field and $\theta$ its dual,
$n$ an odd integer for fermions and even for bosons.
As a consequence the physical observables involve only gradients of $\phi$ or exponential operators
of the field and its dual. It is well known that a massless free field in 1+1 dimensions
has a propagator which is logarithmically growing with distance but here we see that this
pathological behavior does not arises in the observables. All correlation functions are well-behaved and decay
with distance. If the effective field theory for the field $\phi$ is free and massless, then the correlation functions
needed to evaluate physical observables of the microscopic theory can be computed easily.
Now in general the effective theory is never free. In fact the good way to think about this limiting case
is in terms of fixed points in the renormalization group (RG) language. The appearance of exponentials
of the field $\phi$ means that the \textit{scale} of the field is physically meaningful. In fact correlations
have an algebraic decay with exponents that depend upon this scale. This is strikingly different from
the 3+1 QFT behavior where the scale of $\phi$ disappear in the LSZ reduction formula and hence the
S-matrix elements do not depend upon it. If we discard interactions then the massless Bose field in 1+1
dimensions is a line of fixed points and the location along the line is given by the overall $\phi$ scale.
It is often written as the Luttinger parameter $K$ in the condensed matter community and is also known
as the radius of the boson in field theory circles. In general the value of the Luttinger parameter cannot be
computed easily. If we do perturbation theory one can get a weak-coupling estimate. For integrable
models it may be known. This is the case of the all-important Lieb-Liniger gas with delta function interactions
which happens to be realized in ultracold atom systems. Finally several numerical techniques may be
used to extract the value of the Luttinger parameter.

If we now translate the microscopic interactions in terms of the effective field theory the nature of the expansion
Eq.(\ref{defphi}) severely constraints the possible operators that are generated. There are powers of the gradient of the field and
also exponentials of $\phi$ or its dual $\theta$. Strictly speaking there are certainly infinitely many operators in the
effective theory but it is very simple to classify allowed operators according to their relevance/irrelevance
in the RG sense even if no exact solution is available. If all allowed operators are irrelevant for example then
the effective theory in the long-distance limit will be the massless Bose field and accordingly there is
algebraic long-range order instead of the true long-range order of higher-dimensional physics.
If there is a relevant operator then strictly speaking the infrared behavior is out of reach of
perturbation theory. However semiclassical reasoning may then give some clue about the physics of the system.
In the simplest sine-Gordon case, there is also an exact solution which can guide us. So we focus on the case
of a single relevant operator. The Euclidean Lagrangian formulation can be written as~:
\begin{equation}
 \label{SGEq}
\mathcal{L}=\frac{1}{2}(\nabla\phi)^2 -\frac{\alpha}{\beta^2 a^2}\cos (\beta\phi).
\end{equation}
Here we have introduced the two dimensionless couplings $\alpha$ and $\beta$ defining the model as well as 
a length scale $a$ which is a short-distance cut-off. The RG flow of this model has been first studied by Kosterlitz and
Thouless~\cite{KT} but it is only later that a fully consistent calculation of the flow appeared~\cite{AGG}.
The cosine operator is relevant if $\beta^2 < 8\pi$, marginal right at $\beta^2=8\pi$. We thus introduce for
convenience the variable $\delta=\beta^2/8\pi-1$ measuring the distance to marginality.
The flow is then given by the coupled equations~:
\begin{eqnarray}
 a\frac{\partial \alpha}{\partial a}&=&2\alpha\delta +\frac{5}{64}\alpha^3 ,\\
a\frac{\partial \delta}{\partial a}&=&\frac{1}{32}\alpha^2-\frac{1}{16}\alpha^2\delta .
\end{eqnarray}
It is important to note that these equations are valid in the sense of a double perturbative expansion~:
both $\alpha$ and $\delta$ should be small. This fact is often overlooked in the condensed-matter literature
where people extrapolate boldly to arbitrary values of $\beta$ far away from the KT point.
This set of flow equations leads to a portrait which is given in Fig.1. The most salient fact is that there is
a separatrix (in blue) that divides the diagram in two parts. In the figure the long distance limit is found by following the arrows.

\section{sine-Gordon theory for gases in a potential well}
\label{KT}
\begin{figure}
\resizebox{0.75\columnwidth}{!}{
\includegraphics{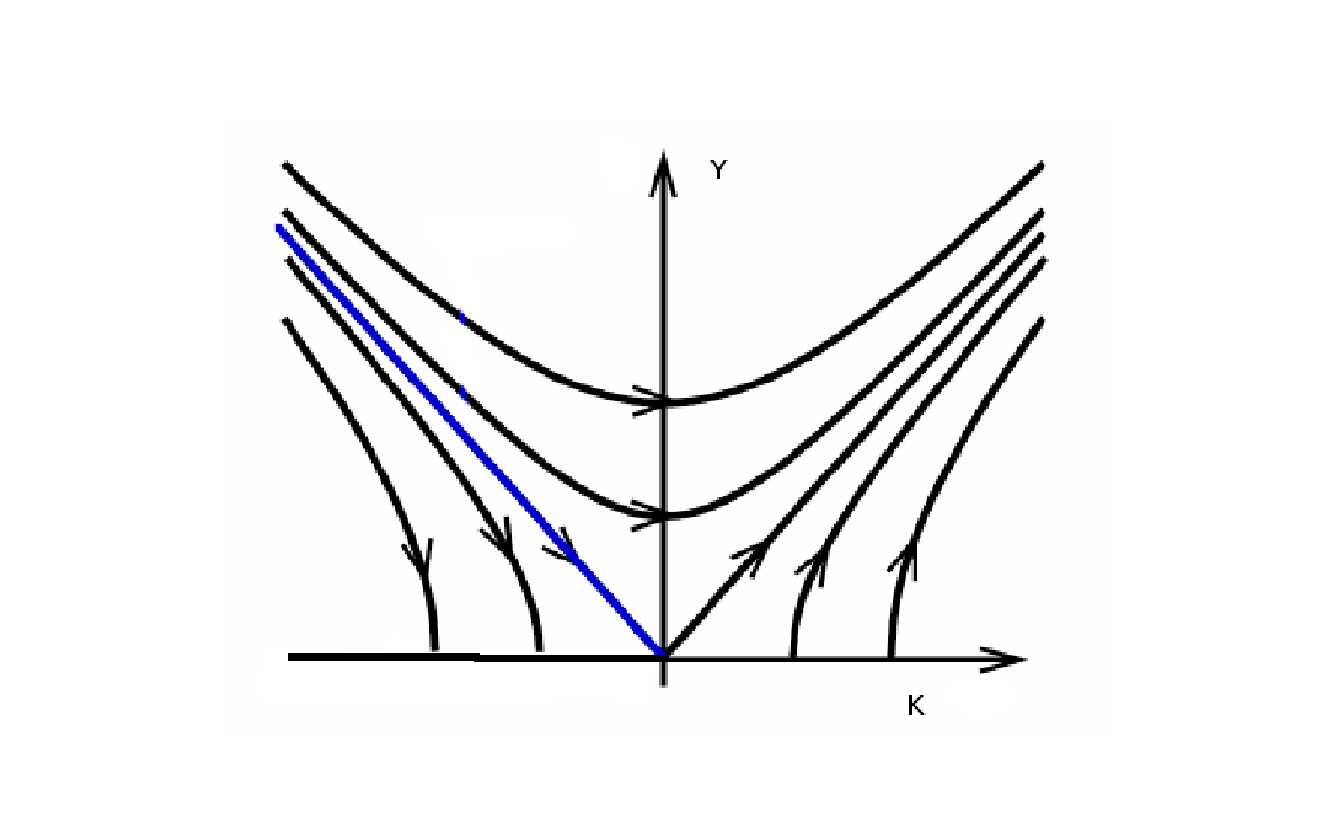} }
\caption{The Kosterlitz-Thouless RG flow~: the vertical axis is the strength of the cosine operator
while the horizontal axis is the Luttinger parameter. This last axis is a line of fixed point which are
stable on the left part and unstable on the right.}
\label{KTflow}       
\end{figure}
Up to now, our discussion has been quite general and we discuss the various physical systems related
to the sine-Gordon flow. The first example is the finite-temperature transition in 2D superfluid systems
like Helium films that was the subject of the original work of Kosterlitz and Thouless.
This transition has been observed in ultracold gases~\cite{ZH06,ZH08}. It is then vortex-unbinding
transition. The vortices have long-range interactions and can be regarded as a 2D Coulomb gas.
It is precisely this problem of statistical mechanics that can be mapped onto a sine-Gordon model~\cite{ChaikinBook}.
If we vary a single parameter, the temperature it means that the equivalent effective model will
move along a line in the 2D figure of the KT flow. When this line crosses the separatrix, the KT transition
takes place. The low-temperature phase being the one with algebraic order in the superfluid phase
and only vortices bound in pairs while above the transition temperature the cosine operator is relevant
and we have unbound vortices. Another physical system which belongs to the same universality class
is the XY spin model where interacting spins are confined in a plane. It is also possible to map this system
onto a 2D Coulomb gas and then onto the sine-Gordon model~\cite{ChaikinBook}.

We now discuss the case of ultracold 1D atomic gases. As mentioned above it is the quantum problem at
T=0K which can be mapped onto the sine-Gordon system. If we have a Bose gas in a weak lattice then
its effective theory is given by the sine-Gordon model. Indeed without an optical lattice an ultracold gas of bosonic atomic is
very precisely described by the Lieb-Liniger model of featureless bosons with delta function interactions that
mimic the ultra-low-energy $s$-wave scattering. For such a system the Luttinger parameter, which is essentially
the coefficient $\beta$ in Eq.(\ref{SGEq}), is related to the dimensionless strength of the interactions $\gamma= mg/n\hbar^2$
where $g$ is the 1D delta function strength and $n$ the density. By using a Feshbach resonance it is possible
to tune the scattering length between the atoms and thus to modify the Luttinger parameter in a controlled way.
Now if we add a weak lattice potential which is commensurate with the density of atoms, this generates in  the
effective theory a cosine operator of the type described in the previous section and its strength is directly proportional
to the strength of the lattice potential. Then the fate of the system depens upon the value of the interactions~\cite{Buchler}. 
If they are weak enough
the cosine operator is irrelevant and the system will remain superfluid for a weak lattice. If we increase the strength
of the potential then at some point the system will cross the separatrix in phase diagram (1) and a KT transition
will take place towards a Mott-Hubbard insulating phase with one boson pinned to each potential well.
This is the standard Mott localization transition. However if we are in the right part of Fig.(1)
then the atoms will be pinned immediately for any arbitrary weak coupling~: the superfluid phase is
immediately destroyed. This phenomenon is expected to be very general and will occur both for Bose and Fermi systems.
It is has been recently demonstrated experimentally by ref.(\cite{Haller2010}). A gas of Cesium atoms
was prepared in an array of 1D tubes and Cs Feshbach resonance allowed the tuning of the interactions.
The gap of the system was measured by an amplitude modulation spectroscopy. The measurements revealing
the immediate development of a Mott state as predicted by the RG treatment of the sine-Gordon model.
In addition there is even quantitative estimates of the gap that are in agreement with the exact results
for the gap of the sine-Gordon model, confirming the validity of the model as well as its phase diagram
with a quantum phase transition.

\section{Two-component systems and trimers}
\label{trimers}
We now turn to the discussion of two-component fermionic systems. These involve mixtures of two ultracold atomic species.
They may be different atoms like mixtures of $^6$Li and $^{40}$K isotopes or even different hyperfine states of a single species when transitions are blocked.
If we put a mixture in an optical lattice it will be described by an asymmetric 
Hubbard model~\cite{Feiguin,GiamarchiCazalillaHo2005,Batrouni2009,DasSarma2009}~:
\begin{equation}
\label{hubbard}
\mathcal{H} = -\sum_{i,\sigma=\uparrow,\downarrow} t_{\sigma} 
[c^{\dag}_{i+1\sigma}c_{i\sigma} + \textrm{h.c.}] + U\sum_i n_{i\uparrow}n_{i\downarrow}\;.
\end{equation}
where the two hoppings are now different $t_\uparrow\neq t_\downarrow$. We consider the case of attractive interactions
$U<0$. Densities are in general different $\rho_\uparrow\neq \rho_\downarrow$ and thus
the Fermi wavevectors are also different~: $k_{F\uparrow}\neq k_{F\downarrow}$.
So it is natural extension of the case of imbalanced superfluids~\cite{FFLO,mit,rice,Orso2007,Drummond2007}.

\begin{figure}
\resizebox{0.6\columnwidth}{!}{
\includegraphics{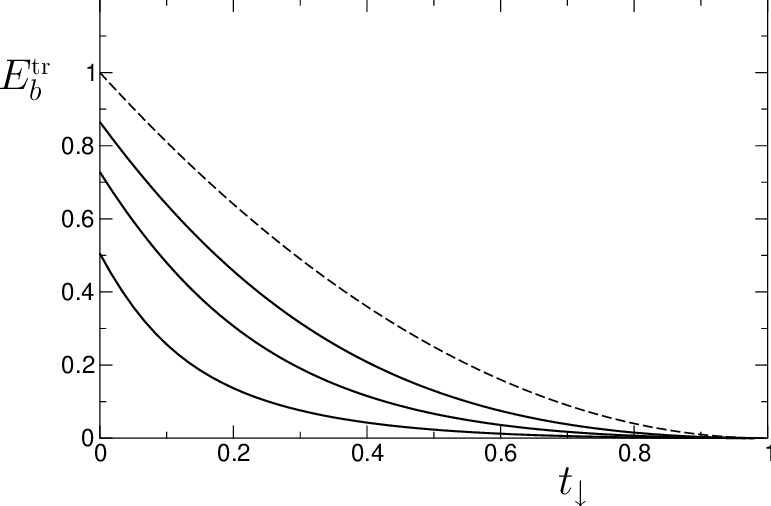} }
\caption{The trimer binding energy as a function of $t_\downarrow$ with $t_\uparrow =1$.
Bottom curve is for $U=-2$ then $-4,-8,-\infty$(dashed line). Note that the trimers always exist.}
\label{binding}       
\end{figure}

With different atoms and attractive interactions one may have few-body bound states. It is possible to study
this phenomenon by looking at the zero-density limit of Eq.(\ref{hubbard}). It is convenient to use the formalism
introduced by D. Mattis~\cite{mattis} that leads to an integral equation whose solution 
allows to determine the bound state spectrum~\cite{trimersL}.
As soon as $t_\uparrow\neq t_\downarrow$, there are three-body bound states for all negative values of $U$.
The trimer binding energy is displayed in fig.(\ref{binding}).
At equal hoppings it is known from the Bethe-Ansatz solution~\cite{takahashi} of the Hubbard model that there are no bound states
with more than two particles so we find vanishing binding in the symmetric limit.

\begin{figure}
\resizebox{0.6\columnwidth}{!}{
\includegraphics{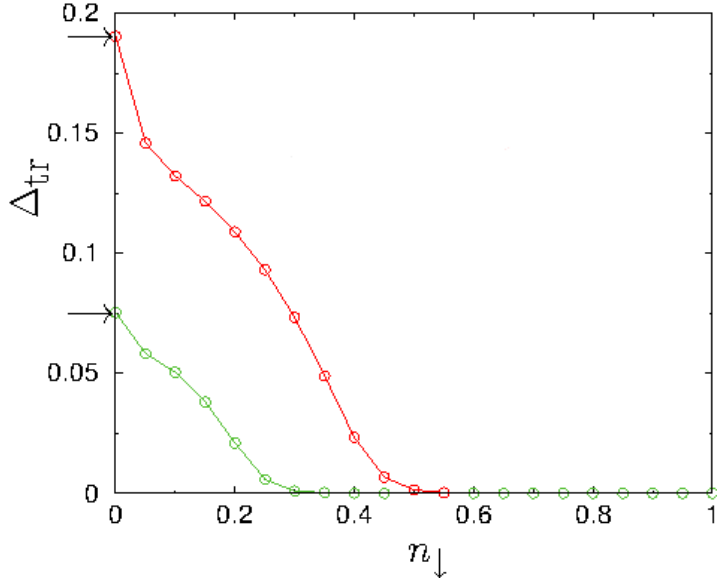} }
\caption{The trimer gap  as a function of $n_\downarrow$ generalizing the binding 
energy at finite density ($n_\downarrow =2n_\uparrow$). The bottom curve is for
$U=-2$ and next curve is for $U=-4$. The zero-density limit is the binding energy
of fig.(\ref{binding} as indicated by arrows.). Note that high density is detrimental
to trimerization.}
\label{binding2}       
\end{figure}

The situation becomes more interesting when we go to finite density. It is not clear that the bound states will survive.
We thus define a many-body trimer gap by~:
\begin{eqnarray}
\label{trimergap}
& \Delta_\mathrm{tr} &=-\lim_{L  \rightarrow \infty} [ E_L(N_\uparrow +1,N_\downarrow+2)
+E_L(N_\uparrow ,N_\downarrow) \nonumber\\
&&-E_L(N_\uparrow +1,N_\downarrow+1)-E_L(N_\uparrow,N_\downarrow+1)]\; ,
\end{eqnarray}
where $ E_L(N_\uparrow,N_\downarrow)$ is the ground state energy of a gas with
spin populations $N_\uparrow,N_\downarrow$ in a chain of size $L$. This quantity is computed by DMRG for large system sizes.
It is plotted in Fig.(\ref{binding2}) where we see that trimers do survive as long as the density is not too high.
It remains to understand the nature of the interacting trimer liquid. We use the bosonization technique applied to the
asymmetric Hubbard model~\cite{bosonization}. There is then a Bose field for each fermionic species $\phi_\uparrow$ and $\phi_\downarrow$.
Each of these fields will have a velocity $v_\sigma$, $\sigma=\uparrow,\downarrow$ as well as a Luttinger parameter
$K_\sigma$. The effective low-energy long-wavelength theory will contain a free part which is the most general
quadratic form in the fields and there are also cosine operators generated by the appearance of higher harmonics
in the expression Eq.(\ref{defphi}). They are typically given by~:
\begin{equation}
\mathcal{O}_{p,q}=
 \cos{ \left[ 2(p k_F^\uparrow - q k_F^\downarrow )x - 2(p\phi_\uparrow - q\phi_\downarrow )  \right]},
\end{equation}
where $p$ and $q$ are integers. Such operators are rapidly oscillating in space as long as
$p\,n_\uparrow - q n_\downarrow \neq0 $ and can be neglected in the continuum limit.
But for special commensurabilities $p\,n_\uparrow - q n_\downarrow \equiv 0 $
we have to consider an operator which survives the continuum limit. 
Note that this effect has nothing to do with the underlying lattice periodicity, it exists even in the continuum.
Due to the appearance of
the integer factor $p$ and $q$ it is not clear that this operator can ever be relevant since its scaling dimension
near the free boson fixed points involves $p^2$ and $q^2$ factors.  In the absence of relevant cosine operators
the quadratic form coupling the two Bose fields can be diagonalized and this leads to two free
eigenmodes~: this is a two-component Luttinger liquid as happens in ordinary condensed-matter systems
involving electrons with spin. Here in the case of an attractive mixture of fermions the physical picture
is a 1D analog of the so-called FFLO phase. All correlations will have algebraic order the pair correlation function
revealing the superconducting properties of the system will oscillate with the FFLO momentum
given by the difference of Fermi wavevectors~: $|k_{F\uparrow}-k_{F\downarrow}|$.

\begin{figure}
\resizebox{0.65\columnwidth}{!}{
\includegraphics{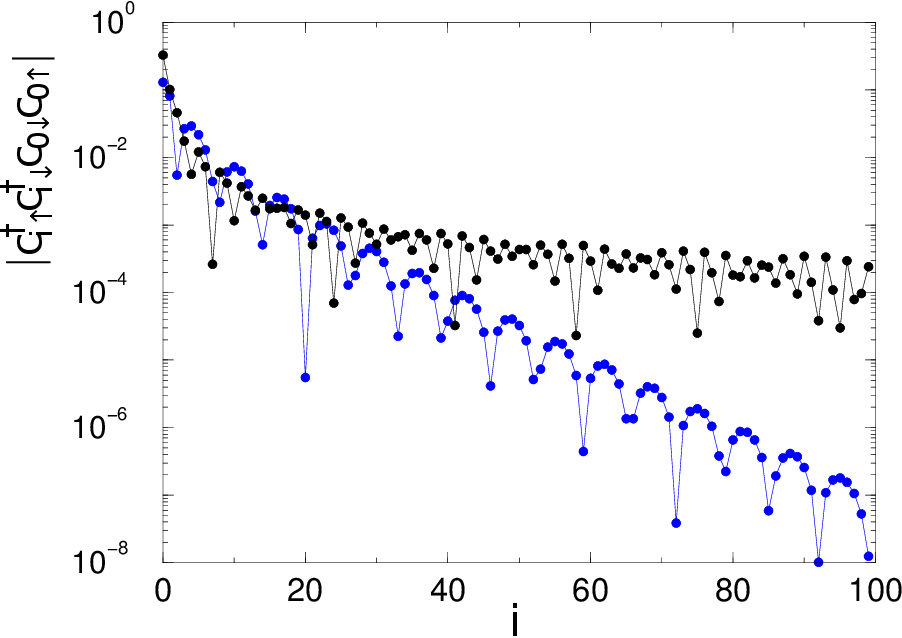} }
\caption{Modulus of superconducting correlations as a function of distance 
We use $U=-4$, $t_\downarrow=0.3$
The upper curve is for $n_\downarrow=0.7, n_\uparrow=0.3$ off-commensurate.
There is algebraic decay. The commensurate case is the lower curve with now
$n_\downarrow=0.6=2n_\uparrow$~: there is exponential decay in the trimer phase.}
\label{corr}       
\end{figure}

If we are at one special commensurability $p\,n_\uparrow - q n_\downarrow = 0 $ for some $p$ and $q$ values  and if the cosine operator is relevant
then a simple semiclassical picture suggests that the cosine will pin the special combination of fields 
$p\phi_\uparrow - q\phi_\downarrow $ to one of the minimum of the potential. This locked mode will have only gapped 
excitations corresponding to small vibration around the minimum energy position while there should exist another
combination of the fields that will remain gapless. This is a one-component Luttinger liquid like the Luther-Emery system.
The question of relevance/irrelevance of this cosine cannot be easily solved by perturbation technique.
A decisive answer can be obtained by using the DMRG~\cite{bosonization,Roux} technique applied to the microscopic model Eq.(\ref{hubbard}).
The ``easiest'' commensurability is expected to be that of 2:1 ratio of densities (otherwise operators are even
less likely to be relevant). Bosonization of correlation functions can be used as a guide to find a way to characterize
the two phases. If we express all correlations in terms of the field which is pinned in the one-component phase
as well as an unknown combination of $\phi_\uparrow$ and $\phi_\downarrow$ which is gapless
then we find that all two-body pairing channels decay exponentially. We thus expect that the simplest microscopic pairing
operator should have the behavior~:

\begin{equation}
\langle c_{x\uparrow}^\dagger c_{x\downarrow}^\dagger
c_{0\downarrow}c_{0\uparrow}\rangle\propto
\frac{\exp(-|x|/\xi)}{|x|^\alpha} \cos(|k_{F\uparrow}-k_{F\downarrow}|x)
\end{equation}

with a correlation length typically $\propto 1/\Delta_\mathrm{tr}$.
So the change from the two-component phase to the one-component phase can be seen by the
change of the decay law of the pair propagator. This is seen in Fig.(\ref{corr}) where we see clearly
that commensurability of type 2:1 leads to a new phase involving locking of the field
$2\phi_\uparrow - \phi_\downarrow$. As is clear from the strong coupling limit, the physical
interpretation is that one has a Luttinger liquid of trimers. This liquid is gapless and has some non-trivial
Luttinger exponent. Excitations corresponding to trimer break-up on the contrary corresponds to gapped modes.
An operator which has algebraic decay in the trimer phase is given simply by the combination
$\Psi_\downarrow^p \Psi_\uparrow^q$. So there is existence of this trimerized phase in the asymmetric Hubbard model.
To map out its domain of existence one can use the fact that the central charge $c$ of the effective theory is either
$c=2$ in the absence of trimers and $c=1$ when they are present~\cite{Roux}. The measurement of the central charge can be done
efficiently by measuring the entanglement entropy by the DMRG algorithm. 

\begin{figure}
\resizebox{0.65\columnwidth}{!}{
\includegraphics{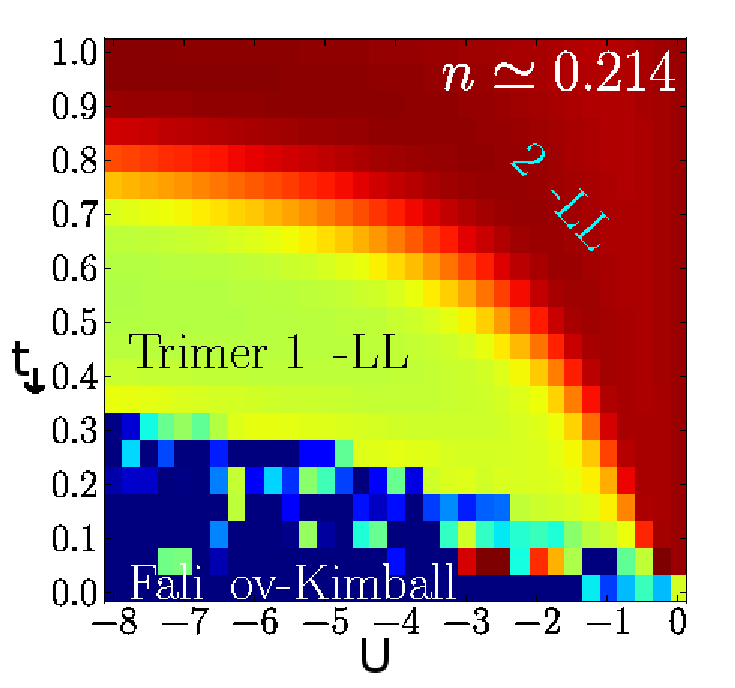} }
\caption{The trimer phase exists for densities not too high and enough hopping
asymmetry. Here the conventional phase is the 2-component Luttinger liquid
and the trimer phase appears for all $U$ by reducing $t_\downarrow$. For higher
densities the trimer phase shrinks and disappears beyond some critical value
$n_{c}$.}
\label{phase}       
\end{figure}

A slice at fixed low density is displayed in Fig.(\ref{phase}). It shows an extended trimer phase for all values of $U$
that exists as soon as the asymmetry is large enough. When the asymmetry becomes too large i.e. $t_\downarrow\rightarrow 0$
we are dealing with the so-called Falicov-Kimball limit. In this case one species does not move and one has to solve
the classical problem of minimizing the energy of the other species. This leads to a complex staircase structure~\cite{Yeomans1997}
which involves all kinds of commensurabilities. This limit however is beyond the reach of bosonization and in addition
the DMRG algorithm no longer converges. The wide region of trimer stability shrinks when we rise the density and finally disappear
for total density close to unity.

\section{Conclusion}

several ultracold gas systems that are under intense scrutiny can be described accurately
by prototypical field theory models like the sine-Gordon model with one or more components.
It is likely that precision measurements will be able to map out phase structure of these systems
in detail as is already the case with the KT system. 
Notably the study of multicomponent systems is hampered theoretically by the fact
that the velocities have no reason to be equal. Thed etailed comparison of theory and experiments
should allow for a better understanding of this complex situation.

Concerning trimers, in principle their detection is accessible to experiments~\cite{castin,esslinger}
and this would allow the experimental observation of these new commensurability effects
which attract much attention~\cite{Tommaso1,Tommaso2}.

\section{Acknowledgments}
One of us (T.J.) would like to thank the organizers of Lyon BEC 2012 to giving the opportunity
to present a short review from which this article is derived.
The detailed DMRG work has been performed in collaboration with G. Roux.


\end{document}